\documentclass[journal=inocaj,manuscript=article]{achemso}
\usepackage[version=3]{mhchem}
\usepackage{xcolor}
\usepackage{siunitx}
\usepackage[hidelinks]{hyperref}
\usepackage{booktabs}
\usepackage{physics}
\usepackage{bm}

\DeclareSIUnit[quantity-product = ]\percent{\char`\%} 
\DeclareSIUnit\angstrom{\text {Å}} 
 \DeclareSIUnit\electronmass{\text {\ensuremath {m}}_{0}}
\newcommand{\hmn}[1]{\ensuremath{\begingroup\setupHMN#1\endgroup}}
\newcommand{\setupHMN}{\doHMN{-}{\HMNoverline}\doHMN{*}{\HMNminverse}\doHMN{i}{\infty}}
\newcommand{\doHMN}[2]{
    \begingroup\lccode`~=`#1
    \lowercase{\endgroup\let~}#2
    \mathcode`#1="8000
}
\newcommand{\HMNminverse}[1]{\frac{#1}{m}}
\newcommand{\HMNoverline}[1]{\mkern1mu\overline{\mkern-1mu#1\mkern-1mu}\mkern1mu}
\newcommand{\bhmn}[1]{\ensuremath{\begingroup\setupHMN\bm{#1}\endgroup}} 

\author{Ruiqi Wu}
\affiliation{Department of Chemistry, Molecular Sciences Research Hub, White City Campus, Imperial College London, Wood Lane, London, UK}

\author{Alex M. Ganose}
\affiliation{Department of Chemistry, Molecular Sciences Research Hub, White City Campus, Imperial College London, Wood Lane, London, UK}
\email{a.ganose@imperial.ac.uk}

\title{Tilt-induced charge localisation in phosphide antiperovskite photovoltaics}

\begin{document}

\begin{abstract}
Antiperovskites are a rich family of compounds with applications in battery cathodes, superconductors, solid-state lighting, and catalysis. 
Recently, a novel series of antimonide phosphide antiperovskites (\ce{$A$3SbP}, where $A$ = Ca, Sr, Ba) were proposed as candidate photovoltaic absorbers due to their ideal band gaps, small effective masses and strong optical absorption.
In this work, we explore this series of compounds in more detail using relativistic hybrid density functional theory.
We reveal that the proposed cubic structures are dynamically unstable and instead identify a tilted orthorhombic \hmn{Pnma} phase as the ground state.
Tilting is shown to induce charge localisation that widens the band gap and increases the effective masses.
Despite this, we demonstrate that the predicted maximum photovoltaic efficiencies remain high (\SIrange{24}{31}{\percent} for \SI{200}{\nm} thin films) by bringing the band gaps into the ideal range for a solar absorber.
Finally, we assess the band alignment of the series and suggest hole and electron contact materials for efficient photovoltaic devices.
\end{abstract}

\section{Introduction}

Semiconductors with the perovskite \ce{$ABX$3} structure (where $A$ and $B$ are cations and $X$ is an anion) have attracted attention as novel photovoltaics, photodetectors, and light-emitting diodes (LEDs).
Reversing the anion and cation sites results in the antiperovskite, or inverse perovskite, \ce{$A$3$YX$} structure where anions occupy the $Y$ and $X$ sites and cations occupy the $A$ sites.
Antiperovskites have been widely studied due to their high ionic conductivity,\cite{kamaya_lithium_2011a} giant magnetoresistance,\cite{kamishima_giant_2000} superconductivity,\cite{bauer_heavy_2004} negative thermal expansion,\cite{chen_negative_2015} tunable luminescent properties,\cite{luo_enhanced_2012} and catalytic performance.\cite{jia_ni3fen_2016}
For this reason, antiperovskites have found use in a broad range of functional applications including superionics for Li and Na ion batteries,\cite{zhao_superionic_2012} white LEDs,\cite{chen_suppressing_2013} and electrocatalysts.\cite{vaughnii_solution_2014}
There is also considerable interest in antiperovskites as novel photovoltaic materials due to the success of their hybrid perovskite counterparts, which have seen efficiencies rise to \SI{26.1}{\percent} in the last decade.\cite{zhao_inactive_2022}

Although almost half of the elements in the periodic table are stable in the antiperovskite structure\cite{wang_antiperovskites_2020} --- satisfying the necessary requirements of ionic radius, electronegativity, and oxidation state --- only a handful of compounds with properties suitable for photovoltaics have been observed experimentally.
\citet{fabini2019candidate} screened 33,000 known semiconductors from the Inorganic Crystal Structure Database (ICSD) based on their thermodynamic stability, band gap, and optical absorption, and identified over 200 candidates with predicted photovoltaic efficiencies greater than \SI{25}{\percent}.
Among their top predictions were antiperovskite oxides \ce{(Ca{,}Sr)3(Si{,}Ge)O} and nitrides \ce{(Ca{,}Sr{,}Ba)3(Sb{,}Bi)N} with high predicted efficiencies and light effective masses.
Similarly, \citet{kuhar2018high} performed a high-throughput search across previously synthesised inorganic compounds and again identified the hexagonal antiperovskite \ce{Ba3SbN} and cubic antiperovskite \ce{Sr3SbN}.
Preliminary defect calculations on the compounds suggested that intrinsic vacancies were shallow and that both may possess a degree of defect tolerance.
However, the use of a generalised gradient approximation (GGA) density functional, as used in their work, is known to underestimate the depth of defect traps.

A number of studies have computationally screened entire compositional spaces for novel antiperovskite photovoltaics.
To date, nitrides have received the most attention, in part due to the preponderance of experimentally known antiperovskite nitride systems (e.g., \citet{heinselman_thin_2019} successfully synthesised thin films of \ce{Mg3SbN} revealing favourable visible light absorption). 
\citet{mochizuki_theoretical_2020} explored the \ce{$A$3$Y$N} space (where $A$ = Mg, Ca, Sr, Ba and $Y$ = P, As, Sb, Bi) through computational elemental substitution in seven crystal structure prototypes.
Based on their small effective masses and high optical absorption, they identified \ce{Mg3PN} and \ce{Sr3PN} as novel candidates and further highlighted \ce{Ba3SbN} and \ce{Sr3SbN} as known promising photovoltaics.
\ce{Ba3SbN} and \ce{Sr3SbN} were also identified by \citet{kang2022antiperovskite} who explored their optoelectronic properties using GW0, finding band gaps of \SI{1.1}{\eV} and \SI{0.85}{\eV}, respectively.
A follow-up study from the same author investigated the defect physics of \ce{Ba3SbN} using hybrid density functional theory (DFT), which is known to provide increased predictive accuracy compared to semi-local GGA functionals.\cite{kang_native_2023}
In contrast to previous work,\cite{kuhar2018high} they revealed low vacancy and interstitial formation energies that suggests the concentrations of these defects will be high at all accessible synthesis conditions.
However, the relatively small charge-capture cross sections of these defects indicates they may not contribute significantly to non-radiative recombination.

A wide range of quaternary antiperovskites have been screened for single-junction and tandem photovoltaic devices.
\citet{SREEDEVI2022106727} studied the \ce{$A{^\prime}A$2PN} and \ce{$A{^\prime}3A$3P2N2} series (where $A$ = Mg, Ca, Sr, Ba, Zn) using hybrid DFT.
Despite identifying \ce{Ba3Sr3P2N2} based on its ideal band gap, its relatively weak optical absorption (only reaching \SI{e5}{\per\cm} at over \SI{3}{\eV}) precludes its use in commercial devices.
Similarly, \citet{han2021design} screened the series \ce{$A$6N2$XX{^\prime}$} (where \ce{$AA{^\prime}$} = PAs, PSb, AsSb, PBi, AsBi, SbBi) and \ce{$A$6$YY{^\prime}X$2} (where \ce{$YY{^\prime}$} = NP, NAs, PAs, and $X$ = Sb, Bi).
Their work identified 5 promising antimonides based on their thermodynamic stability, optical absorption, and high dielectric constants (expected to efficiently screen charged impurities).
Their most promising candidate, \ce{Ca6N2AsSb}, was investigated in a further study using G0W0+SOC and the Bethe--Salpeter equation (BSE) to confirm its optical performance.\cite{D3CP02025H}
This work reported an ideal band gap \SI{1.22}{\eV} and suggested ideal electron and hole contact materials needed for an efficient photovoltaic device.

Clearly, a wide range of antiperovskite nitrides and antimonides have been studied as potential photovoltaics.
However, the phosphide antiperovskites --- in which phosphorous is coordinated to two $A$ site cations --- have received less attention.
A recent report from \citet{liang_predicting_2022} proposed the \ce{$A$3SbP} series (where $A$ = Ca, Sr, Ba) as a novel class of antiperovskite absorbers, with band gaps from \SIrange{0.98}{1.52}{\eV} and relatively small effective masses.
The stability of the series was investigated using elastic constant criteria and the thermodynamic ``energy above hull'' (found to be \SI{0}{\eV} in all cases, indicating no competing phases with lower energy).
However, this work only reported on the cubic phase of the series, whereas antiperovskites are known to adopt a wide range of structural polytypes, including tilted and other low-dimensional structures.
Accordingly, further work is necessary to establish the stability and optoelectronic properties of this class of materials.

\begin{figure}
\includegraphics[width=\linewidth]{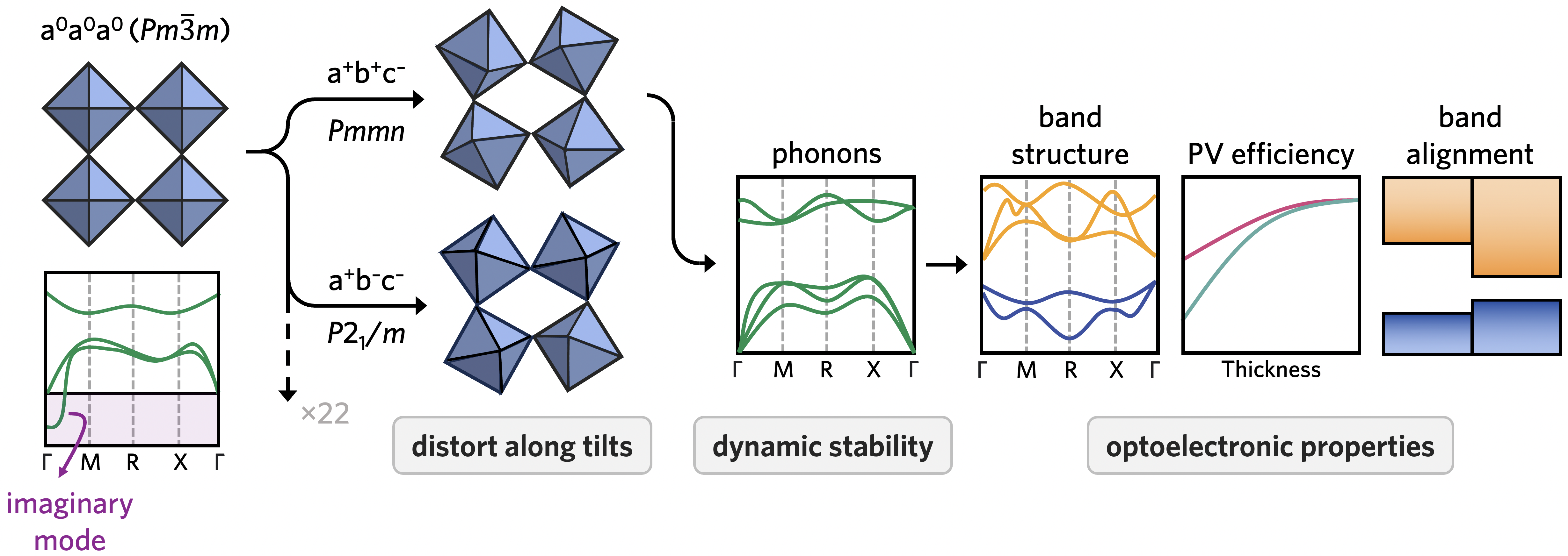}
\caption{Overview of the calculation procedure adopted in this work. The cubic \hmn{Pm-3m} antiperovskite phase of \ce{$A$3SbP} ($A$ = Ca, Sr, Ba) is dynamically unstable, highlighted by imaginary modes in the phonon band structure. To determine the ground state structure, the full set of 22 tilting patterns (indicated by the Glazer tilt notation) are applied and the structures relaxed. The lowest energy structure is selected and the dynamical stability confirmed by the absence of imaginary modes. Finally, the electronic band structure, predicted photovoltaic efficiency, and band alignment are calculated to assess suitability for photovoltaic applications.\label{fig:outline}}
\end{figure}

In this work, we use relativistic hybrid density functional theory to investigate the thermodynamic and dynamical stability, optoelectronic properties, and photovoltaic performance of \ce{$A$3SbP} (where $A$ = Ca, Sr, Ba).
An overview of the computational approach taken is illustrated in Fig.~\ref{fig:outline}.
Vibrational analysis reveals that the cubic \hmn{Pm-3m} phase is dynamically unstable, with imaginary phonon modes indicating spontaneous distortion to lower energy structures.
By exploring the 23 tilting patterns possible in the antiperovskite structure, in addition to a number of low-dimensional polytypes, we identify an orthorhombic \hmn{Pnma} structure as the ground state in all cases.
This tilting is found to induce charge localisation that opens the band gap and slightly reduces the effective masses when compared to the hypothetical cubic structure.
Counterintuitively, these effects only marginally impact the predicted photovoltaic efficiency (which is on the order of \SIrange{24}{31}{\percent}), by bringing the band gaps into the ideal range specified by the detailed-balance limit.
Finally, we assess the band alignment of the series and suggest hole and electron contact materials for efficient photovoltaic devices.

\section{Methodology}

All calculations employed Kohn-Sham density functional theory\cite{kohn_selfconsistent_1965} using the Vienna Ab initio Simulation Package (VASP).\cite{kresse_efficiency_1996,kresse_initio_1993a,kresse_initio_1994a,kresse_ultrasoft_1999}
The plane-wave energy cutoff and $k$-point mesh were converged to a tolerance of \SI{5}{\eV / atom} and \SI{1}{\eV / atom} respectively.
All ground state calculations employed an energy cutoff of \SI{300}{\eV}, with the converged $k$-point meshes presented in Table S1 of the Supplementary Information.
Structural optimisations and vibrational properties were performed using the PBEsol exchange--correlation functional\cite{perdew_restoring_2008}, a version of the PBE functional\cite{perdew_generalized_1996} revised for solids.
The convergence criteria for energy and forces were set to \SI{e-8}{\eV} and \SI{e-2}{\eV\per\angstrom}, respectively.
For accurate descriptions of optoelectronic properties, we employed the Heyd–Scuseria–Ernzerhof hybrid functional (HSE06) for band structures, density of states, and optical absorption calculations.\cite{krukau_influence_2006,heyd_hybrid_2003}
The inclusion of spin--orbit coupling (SOC) effects was essential to describe optoelectronic properties accurately due to the presence of heavy Sb atoms.
This combination of HSE+SOC has been shown to provide a reliable description of the electronic properties for a range of antimony \cite{wang_lone_2022,park_high_2020} and bismuth \cite{ganose_defect_2018} containing materials.
Density of states, band structures, and band edge effective masses were obtained using the \textsc{sumo} package.\cite{alex_sumo_2018}

The high-frequency dielectric response was calculated using HSE+SOC through the frequency-dependent microscopic polarisability matrix as implemented in VASP.\cite{gajdos_linear_2006}
The converged $k$-point meshes used to obtain the dielectric response are presented in Table S1 of the Supplementary Information.
To analyse the upper radiative limit of thickness-dependent energy conversion, we applied the detailed balance method proposed by \citet{blank_selection_2017} --- herein termed the ``Blank'' approach. 
Band alignment calculations employed a slab structure constructed via the \textsc{surfaxe} package\cite{brlec_surfaxe_2021} with a slab and vacuum thickness of \SI{30}{\angstrom} each.
A (100) surface was employed, due to its lack of surface dipole and low surface energy. 

Vibrational properties were calculated within the harmonic approximation using the finite displacement method as implemented in the \textsc{phonopy} package.\cite{togo_firstprinciples_2023}
Vibrational eigenvalues were found to be converged using $3\times3\times3$ and $3\times2\times2$ supercells for the cubic and orthorhombic structures, respectively.
Phonon calculations were performed using the PBEsol functional, with the structure relaxed to a tighter force convergence criterion of \SI{e-4}{\eV\per\angstrom}.
Finite displacements were performed with a \SI{500}{\eV} energy cutoff, with the converged $k$-point meshes presented in Table S1 of the Supplementary Information.

\section{Results and discussion}

\subsection{Thermodynamic and dynamical stability}

\begin{figure}[t]
\includegraphics[width=\linewidth]{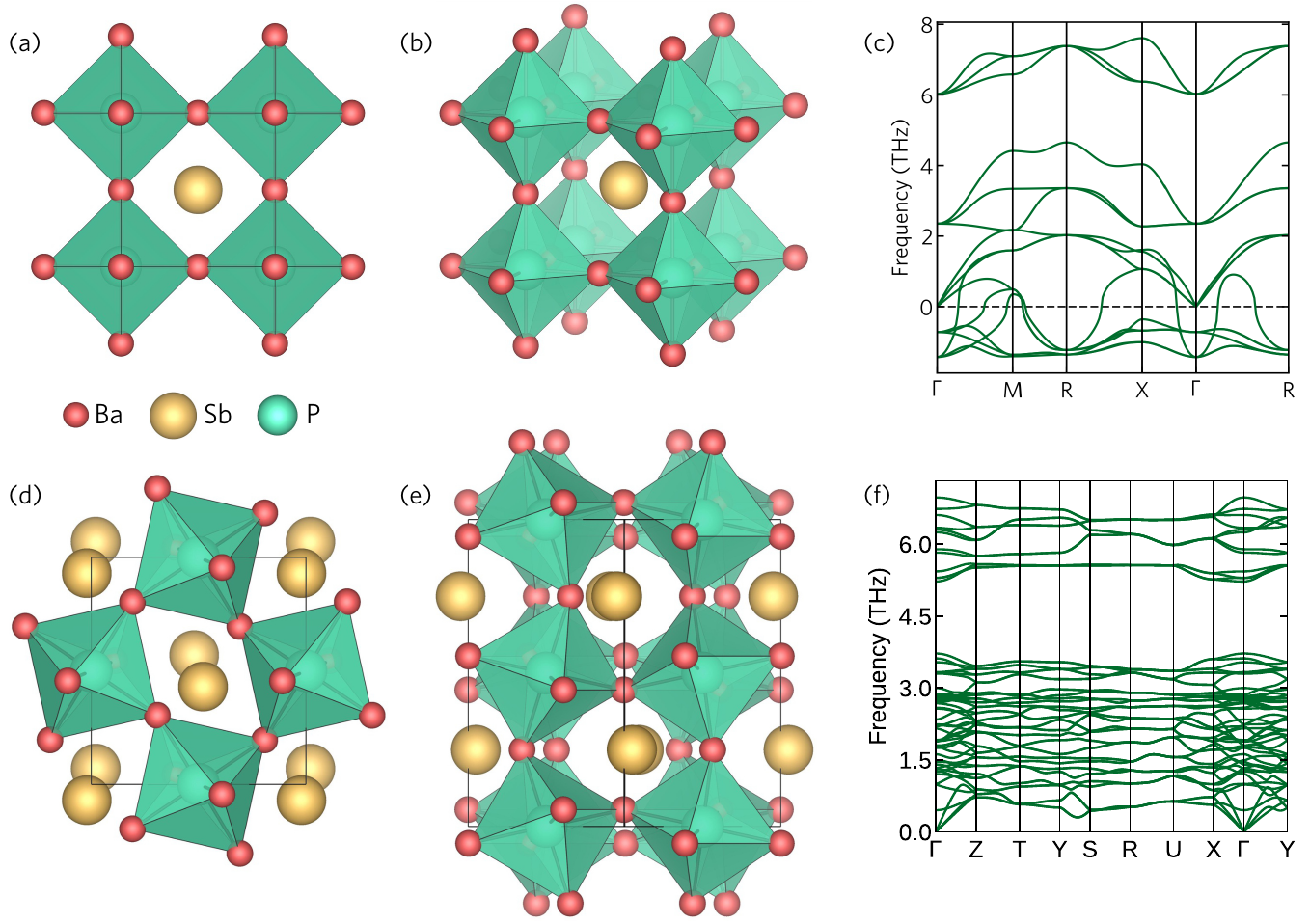}
\caption{(a,b) Crystal structure of cubic \hmn{Pm-3m} and (d,e) orthorhombic \hmn{Pnma} structured \ce{Ba3SbP}, viewed along the (a) [001], (b) free, (c) [010], and (d) [101] directions. The radii of  the spheres has been set to the ionic Shannon radii mentioned in the main text. Phonon band structure of (c) cubic and (d) orthorhombic \ce{Ba3SbP}, calculated using PBEsol.}
\label{fig:structure}
\end{figure}

\citet{liang_predicting_2022} proposed that \ce{$A$3SbP} ($A$ = Ca, Sr, Ba) adopt the cubic antiperovskite structure containing edge sharing \ce{P$A$6} octahedra with Sb in the 12-fold cuboctahedral environment.
In the cubic \hmn{Pm-3m} structure, all octahedra are aligned with no tilting present, resulting in a 5-atom primitive cell (Fig.~\ref{fig:structure}a).
The structural stability of antiperovskites can be predicted using the Goldschmidt tolerance factor,\cite{goldschmidt_gesetze_1926} $t$, as 
\begin{equation}
    t = \frac{r_Y + r_A}{\sqrt{2} (r_X + r_A)},
\end{equation}
where $r_A$, $r_Y$, and $r_X$ correspond to the radii of the $A$, $Y$, and $X$ ions of the \ce{$A$3$YX$} structure.
A tolerance factor greater than 1 indicates a hexagonal or tetragonal perovskite structure is most likely to form, a value between \numrange{0.9}{1} indicates a cubic perovskite, between \numrange{0.71}{0.9} an orthorhombic or rhombohedral perovskite, and a value less than 0.71 is an indicator that alternative non-perovskite structures will be more stable.
We calculate tolerance factors between \numrange{0.77}{0.78} for the \ce{$A$3SbP} series, strongly indicating that a cubic structure is unlikely.
This contrasts with tolerance factors of \numrange{0.89}{0.92} for the nitride antiperovskites \ce{$A$3SbN} (where $A$ = Ca, Sr, Ba), which are known to crystallise in the cubic (Ca, Sr) and hexagonal (Ba) perovskite structures.\cite{mochizuki_theoretical_2020}
We note, typically Goldschmidt tolerance factors are calculated using Shannon's ionic radii, however, the radii for anionic pnictogens \ce{P^{3-}} and \ce{Sb^{3-}} were not reported.
Instead, we use the values derived by \citet{mochizuki_theoretical_2020} --- namely P (\SI{1.82}{\angstrom}) and Sb (\SI{2.12}{\angstrom}) --- with the remaining values taken from Shannon's tabulation as Ca (\SI{1.14}{\angstrom}), Sr (\SI{1.32}{\angstrom}), and Ba (\SI{1.49}{\angstrom}).\cite{shannon_revised_1976}

The instability of the cubic antiperovskite structures is further highlighted through vibrational analysis.
The cubic structures of \ce{$A$3SbP} were relaxed using the PBEsol exchange--correlation functional and the phonons calculated using the finite-displacement approach implemented in \textsc{phonopy}.
The phonon band structures reveal all materials are dynamically unstable with multiple imaginary modes appearing at the $\Gamma$-point and persisting across the Brillouin zone.
The phonon band structure of cubic \ce{Ba3SbP} is illustrated in Fig.~\ref{fig:structure}c as a representative example.
Accordingly, the proposed cubic structures are not a reliable description of the geometry of these compounds, and therefore should not be used to evaluate their photovoltaic performance.

One route to resolving the ground state structure of dynamically unstable phases is through phonon mode-mapping.
In this approach, the structure is displaced along the eigenvectors of an imaginary mode and allowed to relaxed.
The phonons are recalculated and the process repeated until no imaginary modes remain.
This approach has been successfully applied to discover the complex ground state of bismuth stannate containing 176 atoms.\cite{rahim_polymorph_2020}
In contrast, the distortion modes of perovskites were well characterised by Glazer in the 1970s, with only 22 tilting patterns commensurate with the perovskite structure.
These tilting patterns can be expressed in Glazer notation as a$^x$b$^y$c$^z$, where a, b, and c represent the unit cell parameters (if two parameters are the same they share the same letter), while $x$, $y$, $z$ denote the rotation of octahedra along that axis and can take values of 0, $-$, and $+$, indicating no-tilt, anti-phase tilt, and in-phase tilt, respectively.
Accordingly, for perovskite structures, it is often more computationally efficient to distort along each of the 22 modes directly, as it avoids the need for repeated phonon calculations.

For each member of the \ce{$A$3SbP} series, we distort the structure along each of the 22 Glazer tilting patterns, with octahedral rotation angles of $\SI{18}{\degree}$, $\SI{22}{\degree}$, and $\SI{31}{\degree}$ along the a, b, and c axes.
The tilted structures were relaxed using PBEsol and the symmetry of the resulting structure was calculated.
In many cases, the structures relaxed away from their initial distortions, resulting in only 11 symmetry inequivalent tilting patterns.
We also tested a number of perovskite-like polytypes including the \ce{YMnO3} (\hmn{P6_3cm}), ilmenite (\hmn{P6_3/mmc}), rhombohedral perovskite (\hmn{R-3}), and hexagonal perovskite (\hmn{R-3c}) structures which cannot be obtained through simple rotations of the antiperovskite structure.
The full results are presented in Sections S2 and S3 of the Supplementary Information.

We find that an orthorhombic perovskite structure (space group \hmn{Pnma} with a$^+$a$^-$a$^-$ tilting) is the ground state for all members of the \ce{$A$3SbP} series.
In each case, the \hmn{Pnma} structure is considerably lower in energy than the cubic \hmn{Pm-3m} phase, with the distortion resulting in an energy stabilisation of \SI{200}{\meV\per atom} (Ca), \SI{149}{\meV\per atom} (Sr), and \SI{108}{\meV\per atom} (Ba).
This \hmn{Pnma} structure is one of the most commonly observed in antiperovskite materials, such as in \ce{Na3OCl}\cite{pham_computational_2018} and \ce{Li3OCl}\cite{chen_anharmonicity_2015}.
The \hmn{Pnma} structure is lower in energy than the next most stable phase (\hmn{Pmmn} with a$^+$b$^+$b$^-$ tilting for all compositions) by \SI{38}{\meV/atom} (Ca), \SI{32}{\meV/atom} (Sr), and \SI{23}{\meV/atom} (Ba).
We note, for Ba this difference is of the order $k_\mathrm{B}T$ at \SI{300}{\kelvin} ($\sim$\SI{25}{\meV}) and thus the \hmn{Pmmn} structure may be thermally accessible at room temperature.
The dynamic stability of the orthorhombic \hmn{Pnma} phase is confirmed by the phonon band structures which reveal the absence of any imaginary modes, as presented in Fig.~\ref{fig:structure}f for \ce{Ba3SbP}.
Accordingly, we continue with the ground state orthorhombic \hmn{Pnma} structure for the remainder of this work.

As expected, the lattice parameters of the \hmn{Pnma} structures are found to increase in line with the ionic radii of the group 2 elements, with an average increase in lattice parameter of \SI{6}{\percent} for Sr and \SI{12}{\percent} for Ba relative to \ce{Ca3SbP}.
We also find the degree of tilting away from the perfectly cubic structure increases down the group, with the average tilt angles increasing from \SI{35}{\degree} (Ca) to \SI{38}{\degree} (Sr) and \SI{40}{\degree} (Ba).
The increase in tilting is in line with the Goldschmidt tolerance factors of the series, which decrease from Ca to Ba (an indication of greater distortion) as detailed above.
An additional element of distortion in the antiperovskite structure is the displacement of the Sb ions (which sit in the voids between octahedra) away from their idealised positions.
We find the displacement distance also increases down the group, from \SI{0.56}{\angstrom} (Ca) to \SI{0.64}{\angstrom} (Sr) and \SI{0.73}{\angstrom} (Ba).
In Table S6 of the Supplementary Materials, we provide the full set of lattice parameters and tilt angles for the series.

\subsection{Electronic properties}

The band gaps of \ce{$A$3SbP} calculated using HSE+SOC are presented in Table \ref{tab:electronic}.
The band gaps decrease down the group from \SI{1.63}{\eV} (Ca) to \SI{1.62}{\eV} (Sr) an \SI{1.43}{\eV} (Ba).
All band gaps are within the ideal range specified by the detailed balance limit ($\sim$\SIrange{1}{1.6}{\eV} for a maximum efficiency $\sim$\SI{30}{\percent}).
The fundamental band gaps are all direct and appear at the $\Gamma$-point in the Brillouin zone as indicated by the band structure of \ce{Ba3SbP} displayed in Fig.~\ref{fig:band-structure}a.
The band structures for Ca and Sr are presented in Fig.~S2 of the Supplementary Information.
We note that the band gaps are noticeably larger than those of the corresponding antiperovskite nitrides \ce{Sr3SbN} (\SI{0.83}{\eV}) and \ce{Ba3SbN} (\SI{1.10}{\eV})\cite{kang2022antiperovskite}, in part due to the greater octahedral tilting present in the phosphide systems.

\begin{figure}[t]
\includegraphics[width=\linewidth]{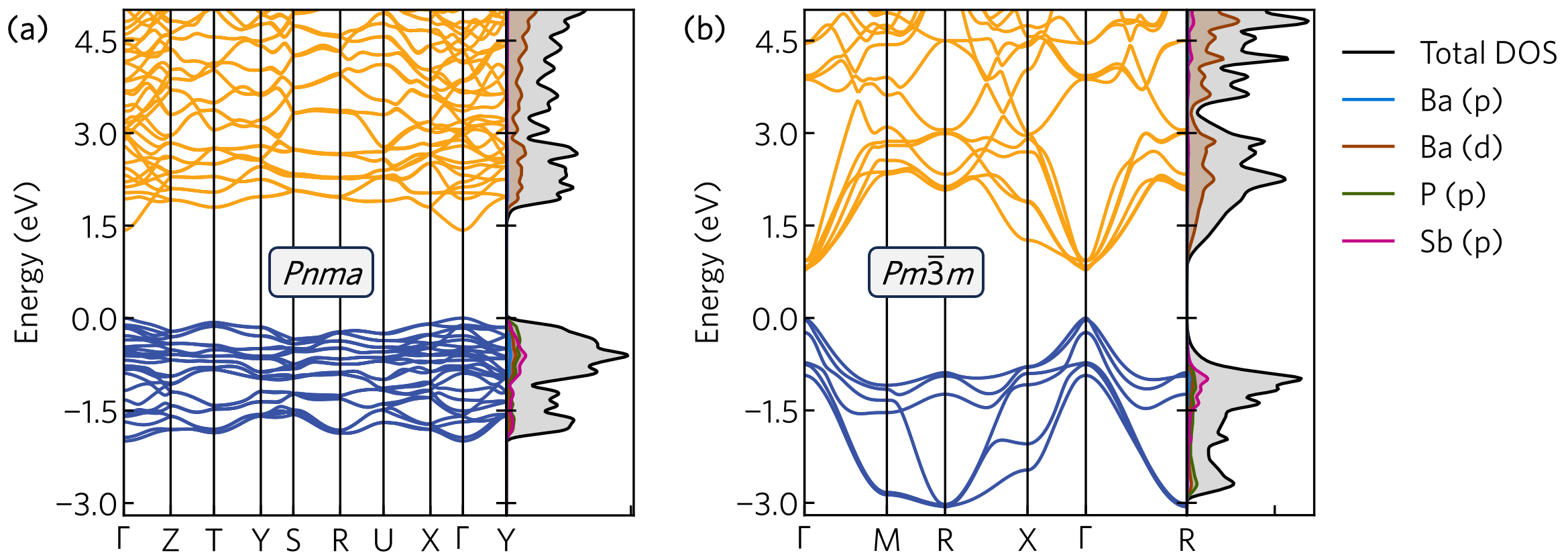}
\caption{Electronic band structure of (a) orthorhombic \hmn{Pnma} and (b) cubic \hmn{Pm-3m} structured \ce{Ba3SbP} calculated using HSE+SOC. The valence band maximum has been set to zero eV.}
\label{fig:band-structure}
\end{figure}

To understand the trend in the band gaps of the series, we examine the orbital contributions at the band edges.
The orbital-projected density of states of \ce{Ba3SbP}, calculated using HSE+SOC is presented in Fig.~\ref{fig:band-structure}a.
Similar to the nitride antiperovskites, the Sb site plays a critical role in the frontier orbitals at the band edges.\cite{kang2022antiperovskite}
This is in contrast to the conventional perovskites, in which the species on the cuboctahedral site are typically spectator ions whose role is to charge balance the composition and provide structural support for the octahedral cages.
The valence band maximum is composed of P 3p and Sb 5p states, whereas the conduction band minimum is almost entirely composed of the Ba 5d orbitals.
The conduction band states are particularly delocalised, as demonstrated by the charge density isosurfaces displayed in Section S4 of the Supplementary Information, and give rise to a relatively disperse band.
As we shall demonstrate later through analysis of the bulk band alignment, the band gap change down the series is largely driven by a shift in the valence band maximum, with the conduction band minimum remaining roughly fixed.
The valence band shift is controlled by two competing factors: i) the bandwidth of the valence band states, which decreases down the series due to the increased bond lengths (from \SI{2.8}{\eV} for Ca to \SI{1.9}{\eV} for Ba) and which acts to open the band gap. ii) The Madelung energies at the Sb and P sites which comprise the valence band edge, which decrease down the series due to the greater octahedral distortions (see Supplementary Table S7 for the tabulated Madelung energies) and which act to lower the band gap.
Finally, we note that spin--orbit coupling has a small but non-negligible impact on the electronic structure, with relativistic renormalisation of \SI{-0.11}{\eV} (Ca), \SI{-0.06}{\eV} (Sr), and \SI{-0.02}{\eV} (Ba).
The diminishing impact of SOC down the group can be attributed to the shrinking contributions of the Sb 5p states --- the only frontier orbitals experiencing strong SOC --- at the valence band edge down the series due to increased tilting (as revealed in the charge density isosurfaces presented in Fig.~S3 of the Supplementary Information).


\begin{table}
\caption{Electronic properties of \ce{$A$3SbP} calculated using HSE+SOC. The band gap ($E_\mathrm{g}$), ionisation potential (IP), and electron affinity (EA) are given in eV. The hole ($m_\mathrm{h}$) and electron ($m_\mathrm{e}$) effective masses are given in units of the electron rest mass, $m_0$. The maximum predicted photovoltaic efficiency for a \SI{200}{\nm} thin film, calculated using the \citet{blank_selection_2017} approach ($\eta_\mathrm{max}$) is given in percent. The effective masses have been averaged using the harmonic mean. The trace of the high-frequency dielectric constants ($\varepsilon_\infty$) has been averaged using the arithmetic mean. The full direction-dependent effective masses and dielectric constants are given in Table S10 of the Supplementary Information}\label{tab:electronic}
\begin{tabular}{@{}lcccccccccccc@{}}
\toprule
\textbf{} & \multicolumn{5}{c}{\bhmn{Pm-3m}} & \multicolumn{7}{c}{\bhmn{Pnma}} \\ \cmidrule(lr){2-6} \cmidrule(l){7-13}
\textbf{Compound} & $\vb*{E}_\mathbf{g}$ & $\vb*{m}_\mathbf{h}$ & $\vb*{m}_\mathbf{e}$ & $\vb*{\varepsilon_\infty}$ & $\vb*{\eta}_\mathbf{max}$ & $\vb*{E}_\mathbf{g}$ & $\vb*{m}_\mathbf{h}$ & $\vb*{m}_\mathbf{e}$ & $\vb*{\varepsilon_\infty}$ & $\vb*{\eta}_\mathbf{max}$ & \textbf{IP} & \textbf{EA} \\ \midrule
\ce{Ca3SbP} & 1.07 & 0.22 & 0.38 & 4.9 & 29.7 & 1.63 & 0.38 & 0.67 & 7.5 & 23.9 & 2.86 & 1.23 \\
\ce{Sr3SbP} & 0.85 & 0.21 & 0.11 & 4.7 & 28.0 & 1.62 & 0.45 & 0.22 & 6.9 & 27.2 & 2.70 & 1.09 \\
\ce{Ba3SbP} & 0.78 & 0.16 & 0.16 & 4.3 & 27.0 & 1.43 & 0.93 & 0.28 & 3.6 & 31.4 & 2.57 & 1.14 \\ \bottomrule
\end{tabular}
\end{table}

To better understand the nature of carrier transport in \ce{$A$3SbP}, we calculate band edge effective masses using the \textsc{sumo} package.\cite{alex_sumo_2018}
The effective masses of the valence band maximum and conduction band minimum are listed in Table \ref{tab:electronic}.
The electron effective masses were found to be relatively light and range from \SI{0.67}{\electronmass} (Ca) to \SI{0.22}{\electronmass} (Sr) and \SI{0.28}{\electronmass} (Ba), as expected based on the dispersive charge density isosurfaces in Fig.~S4 of the Supplementary Information.
This is comparable to the antiperovskite nitrides \ce{Sr3SbN} (\SI{0.22}{\electronmass}) and \ce{Ba3SbN} (\SI{0.31}{\electronmass})\cite{kang2022antiperovskite}, and other more established photovoltaic absorbers, such as \ce{CH3NH3PbI3} ($\mathrm{m}_\mathrm{e}$ = \SI{0.15}{m_0})\cite{frost_atomistic_2014} and CdTe (\SI{0.11}{m_0}).
Small effective masses improve the collection efficiency of photoexcited carriers and typically result in large polarons, weakly bound excitons,\cite{walsh_instilling_2017} and small defect capture cross sections.\cite{huang_highperformance_2022}.
The hole effective masses are somewhat larger, ranging from \SIrange{0.38}{0.93}{\electronmass} across the series.
We note that the effective masses are reasonably anisotropic, particularly for holes, as demonstrated in Table S10 of the Supplementary Information.

We find that the orthorhombic \hmn{Pnma} phase undergoes significant charge localisation compared to the cubic \hmn{Pm-3m} structure.
Primarily, this can be seen in the bandwidth of the upper valence band which reduces from over \SI{3}{\eV} in the cubic phase to $\sim$\SI{1.8}{\eV} in the orthorhombic phase for \ce{Ba3SbP} (Fig.~\ref{fig:band-structure}).
A similar effect can be observed in the conduction band and persists for the other compounds in the series.
This localisation widens the band gap from \SI{0.78}{\eV} to \SI{1.43}{\eV} and increases the electron effective masses from \SI{0.16}{\electronmass} to \SI{0.28}{\electronmass} in the case of the Ba analogue.
This effect is well known in the conventional and hybrid perovskites and originates from increased octahedral tilting that decreases the overlap between the orbitals comprising the band edges.\cite{doi:10.1021/acs.jpclett.0c00295}
We note, however, that this localisation acts to increase the band gaps of the series into the ideal range for efficient solar absorbers (\SIrange{1}{1.6}{\eV}) and thus may be beneficial for photovoltaic device performance.

\subsection{Optical absorption and predicted photovoltaic efficiency}

\begin{figure}
\includegraphics[width=.8\linewidth]{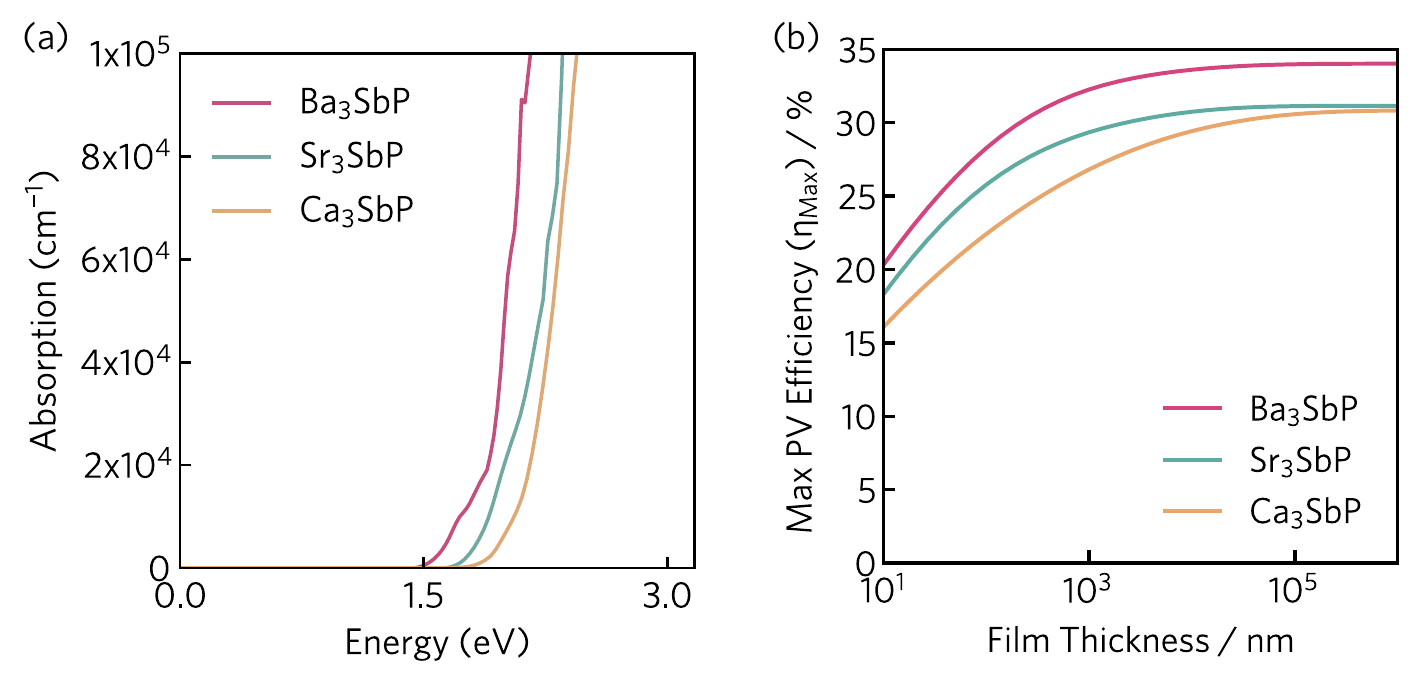}
\caption{a) Optical absorption and b) maximum theoretical efficiency of orthorhombic \hmn{Pnma} structured \ce{$A$3SbP}. Optical absorption was calculated using HSE+SOC. The theoretical efficiency was obtained through the detailed-balance approach developed by \citet{blank_selection_2017}\label{fig:optical}}
\end{figure}

Beyond an appropriate band gap and light charge carriers, strong visible light absorption is essential for efficient photovoltaics.
The optical absorption of \ce{$A$3SbP} calculated using HSE+SOC is presented in Fig.~\ref{fig:optical}a.
All three materials exhibit a steep absorption onset, quickly reaching greater than \SI{e5}{\per\cm} within less than \SI{1}{\eV} of the fundamental band gap.
The strength of the absorption can be attributed to the double degeneracy of the highest valence band and lowest conduction band.
\ce{Ba3SbP} displays the strongest absorption onset, in part due to the more localised valence band edge which increases the joint density of states.
In contrast, \ce{Ca3SbP} exhibits the slowest absorption onset due to its disperse valence band.
The absorption for the series is competitive with other thin-film absorbers such as \ce{CH3NH3PbI3} and CdTe, and is considerably stronger silicon due to its indirect band gap.\cite{dewolf_organometallic_2014}

To assess the potential of the \ce{$A$3SbP} series in photovoltaic devices, we calculate the maximum theoretical efficiency using the detailed balance metric developed by \citet{blank_selection_2017}
This approach goes beyond the standard Shockley--Queisser limit\cite{shockley_detailed_2004} by taking into account the full optical response and distinguishing external luminescence yield from internal radiative recombination.
We model a Lambertian surface that reduces losses through a diffusely scattering front surface and a back surface with unity reflectance.
The theoretical efficiencies were obtained using the frequency-dependent dielectric response calculated using HSE+SOC.
The theoretical efficiency strongly depends on the thickness of the absorber layer and reaches a maximum value of \SI{30}{\percent} (Ca, Sr) and \SI{35}{\percent} (Ba) in the high thickness limit.
For \SI{200}{\nm} films (as commonly used in thin-film photovoltaics), the efficiency is found to be \SI{24}{\percent} (Ca), \SI{27}{\percent} (Sr), and \SI{29}{\percent} (Ba).
This is noticeably larger than predicted for \ce{Sr3SbN} and \ce{Ba3SbN} (\SI{20}{\percent} and \SI{25}{\percent}, respectively\cite{kang2022antiperovskite}), and competitive with other emerging antimony absorbers such as \ce{Sn2SbS2I3} (\SI{29}{\percent}).\cite{kavanagh_hidden_2021}
It is important to stress that the Blank metric provides an upper limit to the photovoltaic performance.
In practical devices, the efficiency will be reduced through non-radiative recombination and contact losses at interfaces.
Regardless, our results highlight the potential of \ce{$A$3SbP} as photovoltaic absorbers.

\subsection{Band alignment}

\begin{figure}
\includegraphics[width=.8\linewidth]{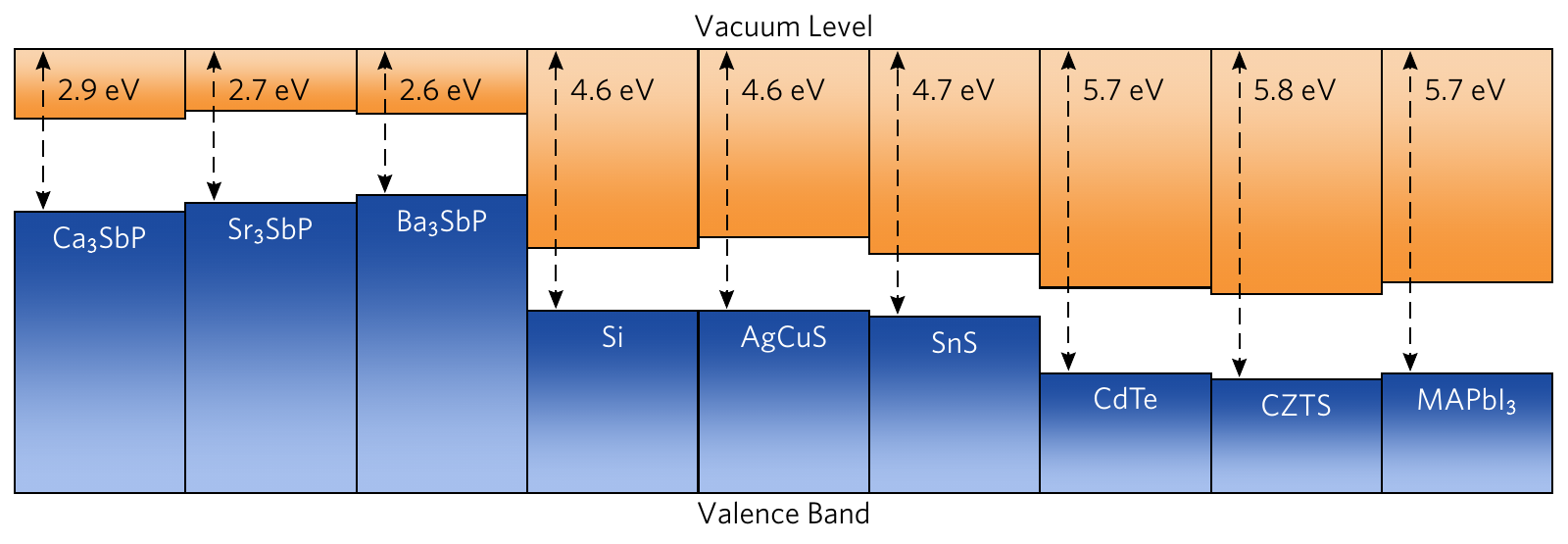}
\caption{Bulk band alignment of \ce{$A$3SbP} calculated using HSE+SOC compared to conventional and emerging photovoltaic absorbers. Ionisation potential and electron affinity of alternate absorbers were taken from Refs.~\citenum{burton_band_2013,brgoch_initio_2014,savory_assessment_2016,walsh_principles_2015}\label{fig:alignment}.}
\end{figure}

Photovoltaic devices with high performance require electron and hole contact materials with suitable alignments to the valence and conduction bands of the absorber.
An efficient alignment prevents open-circuit voltage losses and minimises barriers for carrier diffusion at the interfaces.
To provide insight into ideal device architectures, we perform band alignment calculations using a slab--vacuum model.
Our calculations aim to understand the ``bulk'' band alignment --- namely, they do not take into account relaxations and band bending at the surface.
We calculate ultra-low ionisation potentials and electron affinities of \SIrange{2.6}{2.9}{\eV} and \SIrange{1.1}{1.2}{\eV}, respectively.
We find the depth of the conduction band remains roughly fixed down the series, whereas the valence band edge is raised monotonically from Ca to Ba.
Our calculated band offsets are roughly \SI{1}{\eV} smaller than those found in the nitride antiperovskites.\cite{kang2022antiperovskite}
For example, \citet{kang2022antiperovskite} calculated an ionisation potential and electron affinity of \SI{3.3}{\eV} and \SI{2.4}{\eV} for \ce{Sr3SbN}, using a slab--vaccuum approach with PBE+SOC.
The origin of this shift can be attributed to the greater electronegativity of nitrogen, which acts to push the valence band further from the vacuum level.
In Fig.~\ref{fig:alignment}, we compare the band alignment of the series to a range of conventional and emerging photovoltaic absorbers.
We find the ionisation potentials of \ce{$A$3SbP} to be considerably smaller than any other commonly used absorber, for example Si and AgCuS which are closest in magnitude at \SI{4.6}{\eV}.

The shallow ionisation potentials and electron affinities will require careful choice of contact materials to maximise the obtainable open circuit voltage.
Efficient devices require the work function ($\Phi$) of the transparent electron contact to be greater than the electron affinity of the absorber.
Conversely, the workfunction of the hole transport material should be smaller than the ionisation potential of the absorber.
Most $n$-type transparent conductors have deep workfunctions (greater than \SI{4}{\eV}) and therefore will not be suitable for the phosphide antiperovskites.
Accordingly, ultra-low workfunction electron contact materials such as the semiconducting polymer TPHP(\ce{LiClO4}) ($\Phi$ = \SI{2.2}{\eV}) should be employed.\cite{tong_polymer_2022}
For hole extraction, a low workfunction metal alloy such as MgNd ($\Phi$ = \SI{3.5}{\eV}) may be appropriate.\cite{ge_low_2018}

\section{Conclusion}

We evaluated the \ce{$A$3SbP} antiperovskites (where A = Ca, Sr, Ba) as potential photovoltaics using relativistic hybrid density functional theory calculations.
In contrast to previous reports,\cite{liang_predicting_2022} we demonstrate the cubic \hmn{Pm-3m} phase is likely unstable and will spontaneously distort to an orthorhombic \hmn{Pnma} structure.
We evaluate the thermodynamic stability of this phase against other antiperovskite-like compositions and reveal that it is dynamically stable, with no imaginary modes found in the phonon band structure.
The series possesses ideal band gaps between \SIrange{1.4}{1.6}{\eV}, with small electron effective masses, and strong visible light absorption.
The predicted maximum theoretical efficiencies of \SIrange{24}{31}{\percent} for \SI{200}{\nm} thin films are competitive with other state-of-the-art and emerging absorbers.
Furthermore, we performed band alignment calculations to recommend hole and electron contact materials that will provide efficient charge extraction in optimised devices.
We believe that the \ce{$A$3SbP} series show potential as photovoltaic absorbers and should be targeted for experimental verification.

\section{Author Contributions}

CRediT system for author contributions: https://credit.niso.org
Ruiqi Wu: Investigation, Formal Analysis, Methodology, Visualization, Writing – original draft, Writing – review \& editing.
Alex Ganose: Conceptualization, Methodology, Resources, Supervision, Writing – review \& editing.

\section{Acknowledgements}

A.M.G.~was supported by EPSRC Fellowship EP/T033231/1.
We are grateful to the UK Materials and Molecular Modelling Hub for computational resources, which is partially funded by EPSRC (EP/P020194/1 and EP/T022213/1). 
This work used the ARCHER2 UK National Supercomputing Service (https://www.archer2.ac.uk) via our membership of the UK's HEC Materials Chemistry Consortium, which is funded by EPSRC (EP/L000202).

\bibliography{refs}
\end{document}


\section{Computational methodology}

\begin{table}
\caption{Converged $k$-point meshes used for structural relaxations, optical properties, and finite-displacement phonons of the cubic \hmn{Pm-3m} and orthorhombic \hmn{Pnma} phases of \ce{$A$3SbP}. Density of states calculations used the same $k$-point mesh as optical absorption calculations. All meshes are $\Gamma$-centered}
\begin{tabular}{@{}lcccccc@{}}
\toprule
 & \multicolumn{3}{c}{\bhmn{Pm-3m}} & \multicolumn{3}{c}{\bhmn{Pnma}} \\ \cmidrule(rl){2-4} \cmidrule(l){5-7}
\textbf{Compound} & \textbf{Relax} & \textbf{Optics} & \textbf{Phonons} & \textbf{Relax} & \textbf{Optics} & \textbf{Phonons} \\ \midrule
\ce{Ca3SbP} & $6\times6\times6$ & $6\times6\times6$ & $1\times1\times1$ & $4\times3\times4$ & $5\times4\times5$ & $2\times2\times2$ \\
\ce{Sr3SbP} & $6\times6\times6$ & $6\times6\times6$ & $1\times1\times1$ & $4\times3\times4$ & $6\times5\times7$ & $2\times2\times2$ \\
\ce{Ba3SbP} & $5\times5\times5$ & $5\times5\times5$ & $1\times1\times1$ & $5\times5\times5$ & $6\times5\times7$ & $1\times2\times2$ \\
\bottomrule
\end{tabular}
\end{table}

\clearpage
\section{Energetics of antiperovskite tilted phases}

\begin{table}
\caption{Summary of titling Glazer notation, space group before and after relaxation, and normalized energy with respect to cubic structure calculated by hybrid functional HSE06 with SOC for \ce{Ba3SbP}}
\begin{tabular}{@{}cccc@{}}
\toprule
 & \multicolumn{2}{c}{\textbf{Space group}} & \textbf{} \\ \cmidrule(lr){2-3}
\textbf{Glazer Notation} & \textbf{Initial} & \textbf{Relaxed} & $\mathbf{\Delta}\vb*{E}$ \textbf{(meV/atom)} \\ \midrule
a$^+$b$^+$c$^+$ & \hmn{Immm} (71) & \hmn{I4/mmm} (139) & -132 \\
a$^+$b$^+$b$^+$ & \hmn{Immm2} (71) & \hmn{I4/mmm} (139) & -132 \\
a$^+$a$^+$a$^+$ & \hmn{Im-3} (204) & \hmn{Im-3} (204) & -128 \\
a$^+$b$^+$c$^-$ & \hmn{Pmmn} (59) & \hmn{Pmmn} (59) & -162 \\
a$^+$a$^+$c$^-$ & \hmn{P4_2/nmc} (137) & \hmn{P4_2/nmc} (137) & -151 \\
a$^+$b$^+$b$^-$ & \hmn{Pmmn} (59) & \hmn{Pmmn} (59) & -162 \\
a$^+$a$^+$a$^-$ & \hmn{P4_2/nmc} (137) & \hmn{P4_2/nmc} (137) & -151 \\
a$^+$b$^-$c$^-$ & \hmn{P2_1/m} (11) & \hmn{Pnma} (62) & -200 \\
a$^+$a$^-$c$^-$ & \hmn{P2_1/m} (11) & \hmn{Pnma} (62) & -200 \\
a$^+$b$^-$b$^-$ & \hmn{Pnma} (62) & \hmn{Pnma} (62) & -200 \\
\textbf{a$^+$a$^-$a$^-$} & $ \bhmn{Pnma}$ \textbf{(62)} & $ \bhmn{Pnma}$ \textbf{(62)} & \textbf{-200} \\
a$^-$b$^-$c$^-$ & \hmn{P-1} (2) & \hmn{C2/m} (12) & -141 \\
a$^-$b$^-$b$^-$ & \hmn{C2/c} (15) & \hmn{Imma} (74) & -141 \\
a$^-$a$^-$a$^-$ & \hmn{R-3c} (167) & \hmn{R-3c} (167) & -132 \\
a$^0$b$^+$c$^+$ & \hmn{Immm} (71) & \hmn{I4/mmm} (139) & -132 \\
a$^0$b$^+$b$^+$ & \hmn{I4/mmm} (139) & \hmn{I4/mmm} (139) & -132 \\
a$^0$b$^+$c$^-$ & \hmn{Cmcm} (63) & \hmn{Cmcm} (63) & -145 \\
a$^0$b$^+$b$^-$ & \hmn{Cmcm} (63) & \hmn{Cmcm} (63) & -145 \\
a$^0$b$^-$c$^-$ & \hmn{C2/m} (12) & \hmn{C2/m} (12) & -141 \\
a$^0$b$^-$b$^-$ & \hmn{Imma} (74) & \hmn{Imma} (74) & -141 \\
a$^0$a$^0$c$^+$ & \hmn{P4/mbm} (127) & \hmn{P4/mbm} (127) & -118 \\
a$^0$a$^0$c$^-$ & \hmn{I4/mmm} (139) & \hmn{I4/mcm} (140) & -113 \\
a$^0$a$^0$a$^0$ & \hmn{Pm-3m} (221) & \hmn{Pm-3m} (221) & 0 \\ \bottomrule
\end{tabular}
\end{table}

\begin{table}
\caption{Summary of titling Glazer notation, space group before and after relaxation, and normalized energy with respect to cubic structure calculated by hybrid functional HSE06 with SOC for \ce{Sr3SbP}}
\begin{tabular}{@{}cccc@{}}
\toprule
 & \multicolumn{2}{c}{\textbf{Space group}} & \textbf{} \\ \cmidrule(lr){2-3}
\textbf{Glazer Notation} & \textbf{Initial} & \textbf{Relaxed} & $\mathbf{\Delta}\vb*{E}$ \textbf{(meV/atom)} \\ \midrule
a$^+$b$^+$c$^+$ & \hmn{Immm} (71) & \hmn{Immm} (71) & -97 \\
a$^+$b$^+$b$^+$ & \hmn{Immm2} (71) & \hmn{I4/mmm(139)} & -97 \\
a$^+$a$^+$a$^+$ & \hmn{Im-3} (204) & \hmn{Im-3} (204) & -95 \\
a$^+$b$^+$c$^-$ & \hmn{Pmmn} (59) & \hmn{Pmmn} (59) & -117 \\
a$^+$a$^+$c$^-$ & \hmn{P4_2/nmc} (137) & \hmn{P4_2/nmc} (137) & -111 \\
a$^+$b$^+$b$^-$ & \hmn{Pmmn} (59) & \hmn{Pmmn} (59) & -117 \\
a$^+$a$^+$a$^-$ & \hmn{P4_2/nmc} (137) & \hmn{P4_2/nmc} (137) & -111 \\
a$^+$b$^-$c$^-$ & \hmn{P2_1/m} (11) & \hmn{Pnma} (62) & -149 \\
a$^+$a$^-$c$^-$ & \hmn{P2_1/m} (11) & \hmn{Pnma} (62) & -149 \\
a$^+$b$^-$b$^-$ & \hmn{Pnma} (62) & \hmn{Pnma} (62) & -149 \\
\textbf{a$^+$a$^-$a$^-$} & \bhmn{Pnma} \textbf{(62)} & \bhmn{Pnma} \textbf{(62)} & \textbf{-149} \\
a$^-$b$^-$c$^-$ & \hmn{P-1} (2) & \hmn{Imma} (74) & -102 \\
a$^-$b$^-$b$^-$ & \hmn{C2/c} (15) & \hmn{Imma} (74) & -102 \\
a$^-$a$^-$a$^-$ & \hmn{R-3c} (167) & \hmn{R-3c} (167) & -94 \\
a$^0$b$^+$c$^+$ & \hmn{Immm} (71) & \hmn{Immm} (71) & -97 \\
a$^0$b$^+$b$^+$ & \hmn{I4/mmm} (139) & \hmn{I4/mmm} (139) & -97 \\
a$^0$b$^+$c$^-$ & \hmn{Cmcm} (63) & \hmn{Cmcm} (63) & -108 \\
a$^0$b$^+$b$^-$ & \hmn{Cmcm} (63) & \hmn{Cmcm} (63) & -108 \\
a$^0$b$^-$c$^-$ & \hmn{C2/m} (12) & \hmn{Imma} (74) & -102 \\
a$^0$b$^-$b$^-$ & \hmn{Imma} (74) & \hmn{Imma} (74) & -102 \\
a$^0$a$^0$c$^+$ & \hmn{P4/mbm} (127) & \hmn{Pm-3m} (221) & 0 \\
a$^0$a$^0$c$^-$ & \hmn{I4/mmm} (139) & \hmn{Ima2} (46) & -105 \\
a$^0$a$^0$a$^0$ & \hmn{Pm-3m} (221) & \hmn{Pm-3m} (221) & 0 \\\bottomrule
\end{tabular}
\end{table}

\begin{table}
\caption{Summary of titling Glazer notation, space group before and after relaxation, and normalized energy with respect to cubic structure calculated by hybrid functional HSE06 with SOC for \ce{Ca3SbP}}
\begin{tabular}{@{}cccc@{}}
\toprule
 & \multicolumn{2}{c}{\textbf{Space group}} & \textbf{} \\ \cmidrule(lr){2-3}
\textbf{Glazer Notation} & \textbf{Initial} & \textbf{Relaxed} & $\mathbf{\Delta}\vb*{E}$ \textbf{(meV/atom)} \\ \midrule
a$^+$b$^+$c$^+$ & \hmn{Immm} (71) & \hmn{Immm} (71) & -73 \\
a$^+$b$^+$b$^+$ & \hmn{Immm2} (71) & \hmn{Immm} (71) & -73 \\
a$^+$a$^+$a$^+$ & \hmn{Im-3} (204) & \hmn{Im-3} (204) & -70 \\
a$^+$b$^+$c$^-$ & \hmn{Pmmn} (59) & \hmn{Pmmn} (59) & -85 \\
a$^+$a$^+$c$^-$ & \hmn{P4_2/nmc} (137) & \hmn{P4_2/nmc} (137) & -81 \\
a$^+$b$^+$b$^-$ & \hmn{Pmmn} (59) & \hmn{Pmmn} (59) & -85 \\
a$^+$a$^+$a$^-$ & \hmn{P4_2/nmc} (137) & \hmn{P4_2/nmc} (137) & -81 \\
a$^+$b$^-$c$^-$ & \hmn{P2_1/m} (11) & \hmn{Pnma} (62) & -108 \\
a$^+$a$^-$c$^-$ & \hmn{P2_1/m} (11) & \hmn{Pnma} (62) & -108 \\
a$^+$b$^-$b$^-$ & \hmn{Pnma} (62) & \hmn{Pnma} (62) & -108 \\
\textbf{a$^+$a$^-$a$^-$} & \bhmn{Pnma} \textbf{(62)} & \bhmn{Pnma} \textbf{(62)} & \textbf{-108} \\
a$^-$b$^-$c$^-$ & \hmn{P-1} (2) & \hmn{C2/m} (12) & -74 \\
a$^-$b$^-$b$^-$ & \hmn{C2/c} (15) & \hmn{Imma} (74) & -74 \\
a$^-$a$^-$a$^-$ & \hmn{R-3c} (167) & \hmn{R-3c} (167) & -68 \\
a$^0$b$^+$c$^+$ & \hmn{Immm} (71) & \hmn{Immm} (71) & -73 \\
a$^0$b$^+$b$^+$ & \hmn{I4/mmm} (139) & \hmn{I4/mmm} (139) & -73 \\
a$^0$b$^+$c$^-$ & \hmn{Cmcm} (63) & \hmn{Cmcm} (63) & -81 \\
a$^0$b$^+$b$^-$ & \hmn{Cmcm} (63) & \hmn{Cmcm} (63) & -81 \\
a$^0$b$^-$c$^-$ & \hmn{C2/m} (12) & \hmn{C2/m} (12) & -74 \\
a$^0$b$^-$b$^-$ & \hmn{Imma} (74) & \hmn{Imma} (74) & -75 \\
a$^0$a$^0$c$^+$ & \hmn{P4/mbm} (127) & \hmn{Pm-3m} (221) & 0 \\
a$^0$a$^0$c$^-$ & \hmn{I4/mmm} (139) & \hmn{I4/mmm} (139) & -73 \\
a$^0$a$^0$a$^0$ & \hmn{Pm-3m} (221) & \hmn{Pm-3m} (221) & 0 \\\bottomrule
\end{tabular}
\end{table}

\clearpage
\section{Energetics of antiperovskite polytypes}

\begin{table}
\caption{Normalized energy per atom of antiperovskite polytype phases compared to the orthorhombic \hmn{Pnma} structure}
\begin{tabular}{lccc}
\toprule
\textbf{Structural polytype} & \textbf{Ca} & \textbf{Sr} & \textbf{Ba} \\ \midrule
\hmn{Pnma} (our structure) & 0 & 0 & 0 \\
\hmn{P6_3cm} (\ce{YMnO3}-type) & 103 & 102 & 119 \\
\hmn{P6_3/mmc} (ilmenite-type) & 164 & 184 & 189 \\
\hmn{R-3} (rhombohedral perovskite) & 86 & 76 & 92 \\
\hmn{R-3c} (hexagonal perovskite) & 39 & 55 & 221 \\ \bottomrule
\end{tabular}
\end{table}

\section{Structural parameters of orthorhombic phases}

\begin{table}
\caption{Geometric parameters of orthorhombic \hmn{Pnma} structures for the \ce{$A$3SbP} series. Lattice parameters, $a$, $b$, and $c$ given in \si{\angstrom}. All cell angles were \SI{90}{\degree}, consistent with the orthorhombic space group. The \hmn{Pnma} structure displays two distinct tilting angles, one perpendicular to the $b$ axis ($\angle_\perp$) and another parallel to the $b$ axis ($\angle_\parallel$), where the untilted cubic phase has tilting angles of \SI{0}{\degree}. The displacement of Sb away from the idealised coordinate, $d_\mathrm{Sb}$ is given in \si{\angstrom}. We also provide the dimensional Goldschmidt tolerance factor, $t$, as detailed in the main text.}
\begin{tabular}{lccccccc}
\toprule
\textbf{Compound} & $\vb*{a}$ & $\vb*{b}$ & $\vb*{c}$ & $\angle_\parallel$ & $\angle_\perp$ & $\vb*{d}_\mathbf{Sb}$ & $\vb*{t}$ \\ \midrule
\ce{Ca3SbP} & 7.752 & 10.483 & 7.331 & 36.9 & 34.6 & 0.56 & 0.7788 \\
\ce{Sr3SbP} & 8.255 & 11.094 & 7.725 & 39.3 & 36.3 & 0.64 & 0.7747 \\
\ce{Ba3SbP} & 8.775 & 11.687 & 8.108 & 41.7 & 37.9 & 0.73 & 0.7712 \\\bottomrule
\end{tabular}
\end{table}

\clearpage
\section{Electronic structure}

\begin{table}
\caption{Site Madelung energies for the orthorhombic \hmn{Pnma} phase of \ce{$A$3SbP}. The Madelung energies are given in \si{\eV}. The letters in parentheses are the Wyckoff labels. Madelung energies were calculated using the Ewald summation approach as implemented in the \textsc{pymatgen} software library.}
\begin{tabular}{lcccc}
\toprule
\textbf{Compound} & $\vb*{A}$ \textbf{(c)} & $\vb*{A}$ \textbf{(d)} & \textbf{Sb (c)} & \textbf{P (a)} \\ \midrule
\ce{Ca3SbP} & -16.89 & -16.61 & -32.72 & -38.80 \\
\ce{Sr3SbP} & -15.96 & -15.71 & -31.11 & -36.28 \\
\ce{Ba3SbP} & -15.11 & -14.91 & -29.68 & -34.02 \\\bottomrule
\end{tabular}
\end{table}

\begin{figure}
\centering
\includegraphics[width=\linewidth]{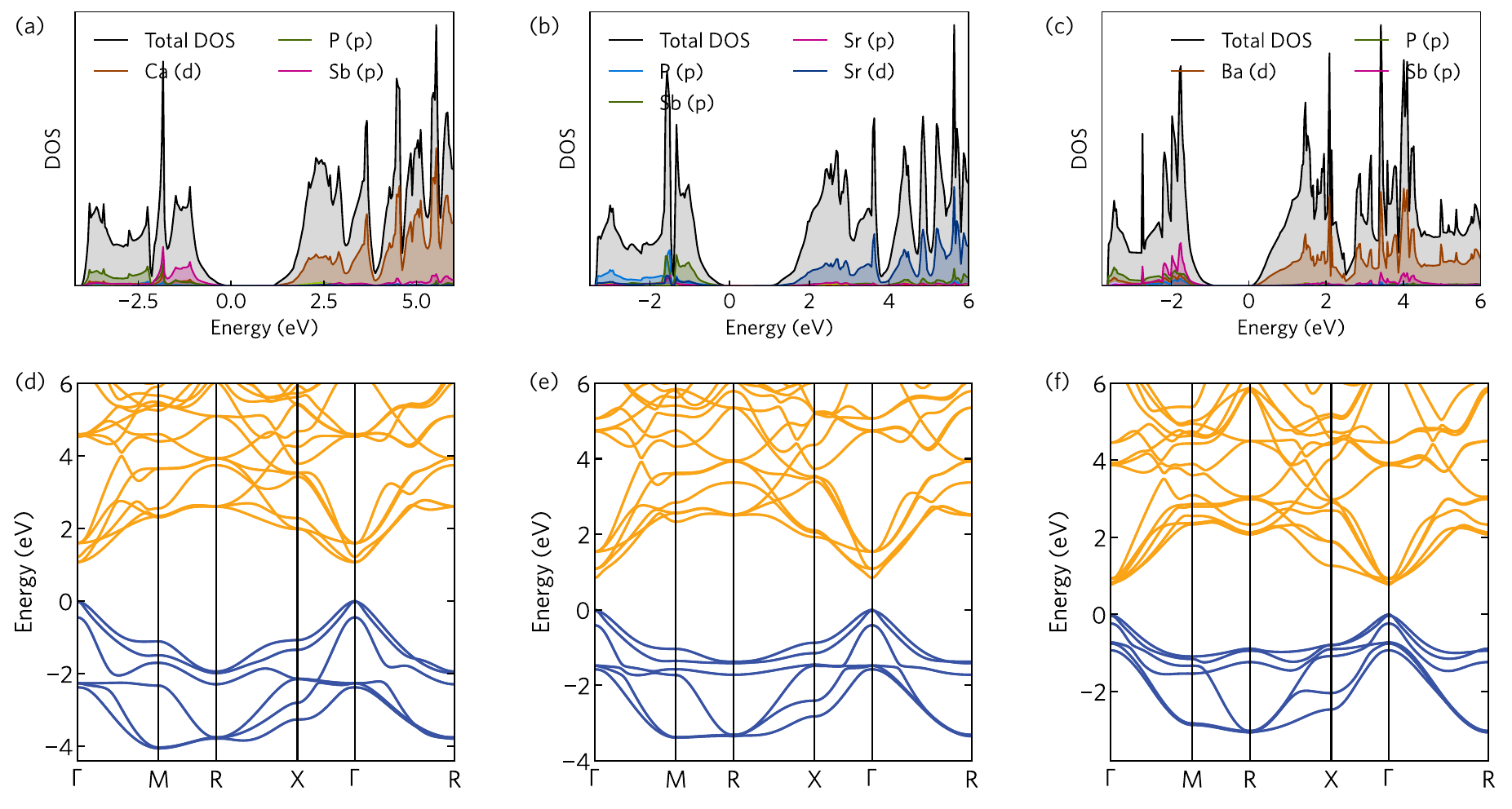}
\caption{Electronic properties of the cubic \hmn{Pm-3m} phase of \ce{$A$3SbP} calculated using HSE06+SOC. (a-c) Density of states of \ce{Ca3SbP}, \ce{Sr3SbP} and \ce{Ba3SbP}; (d-f) Electronic band structure of \ce{Ca3SbP}, \ce{Sr3SbP} and \ce{Ba3SbP}.}
\end{figure}

\begin{figure}
\centering
\includegraphics[width=\linewidth]{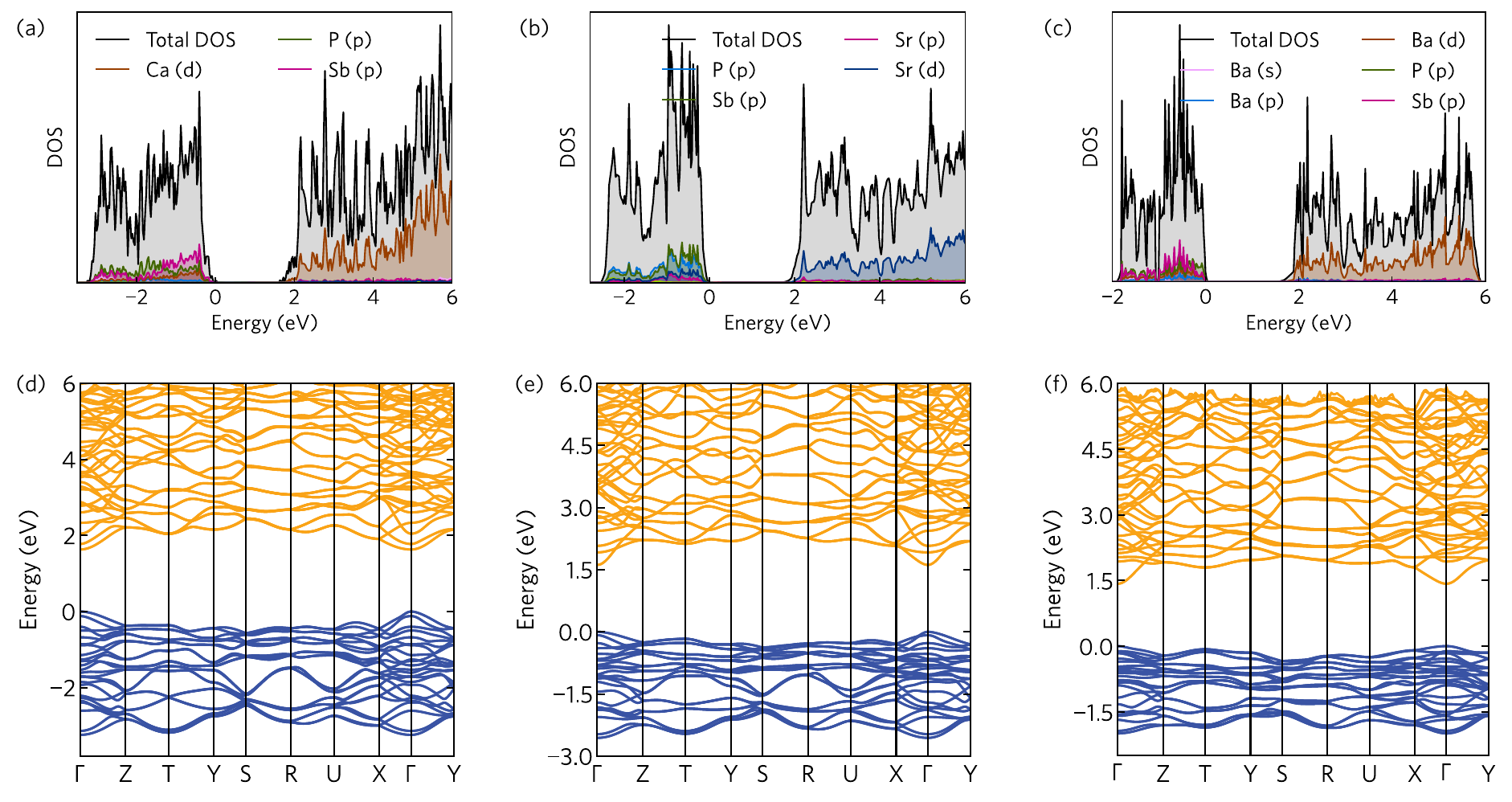}
\caption{Electronic properties of the orthorhombic \hmn{Pnma} phase of \ce{$A$3SbP} calculated using HSE06+SOC. (a-c) Density of states of \ce{Ca3SbP}, \ce{Sr3SbP} and \ce{Ba3SbP}; (d-f) Electronic band structure of \ce{Ca3SbP}, \ce{Sr3SbP} and \ce{Ba3SbP}.}
\end{figure}

\begin{table}
\caption{Band gaps calculated with ($E_\mathrm{g}^\mathrm{soc}$) and without ($E_\mathrm{g}$) spin--orbit coupling (SOC) and the HSE06 functional for the cubic \hmn{Pm-3m} and orthorhombic \hmn{Pnma} phases of \ce{$A$3SbP}. The impact of SOC is given by $\Delta E_\mathrm{soc}$. All values are given in eV}
\begin{tabular}{@{}lcccccc@{}}
\toprule
& \multicolumn{3}{c}{\bhmn{Pm-3m}} & \multicolumn{3}{c}{\bhmn{Pnma}} \\ \cmidrule(lr){2-4} \cmidrule(l){5-7}
\textbf{Compound} & $\vb*{E}_\mathbf{g}$ & $\vb*{E}_\mathbf{g}^\mathbf{soc}$ & $\mathbf{\Delta}\vb*{E}_\mathbf{soc}$ & $\vb*{E}_\mathbf{g}$ & $\vb*{E}_\mathbf{g}^\mathbf{soc}$ & $\mathbf{\Delta}\vb*{E}_\mathbf{soc}$ \\ \midrule
\ce{Ca3SbP} & 1.23 & 1.07 & -0.16 & 1.74 & 1.63 & -0.11 \\
\ce{Sr3SbP} & 1.00 & 0.85 & -0.15 & 1.68 & 1.62 & -0.06 \\
\ce{Ba3SbP} & 0.91 & 0.78 & -0.13 & 1.45 & 1.43 & -0.02 \\ \bottomrule
\end{tabular}
\end{table}

\clearpage
\begin{table}
\caption{Hole ($m_\mathrm{h}$) and electron ($m_\mathrm{e}$) effective masses of the cubic \hmn{Pm-3m} phase of \ce{$A$3SbP}, calculated using HSE+SOC. Effective masses given in units of the electron rest mass, $m_0$}
\begin{tabular}{@{}lcccccccc@{}}
\toprule
 & \multicolumn{4}{c}{$\vb*{m}_\mathbf{h}$} & \multicolumn{4}{c}{$\vb*{m}_\mathbf{e}$} \\ \cmidrule(lr){2-5} \cmidrule(l){6-9} 
\textbf{Compound} & $\mathbf{\Gamma\rightarrow}$\textbf{Z} & $\mathbf{\Gamma\rightarrow}$\textbf{X} & $\mathbf{\Gamma\rightarrow}$\textbf{Y} & \textbf{Avg.} & $\mathbf{\Gamma\rightarrow}$\textbf{Z} & $\mathbf{\Gamma\rightarrow}$\textbf{X} & $\mathbf{\Gamma\rightarrow}$\textbf{Y} & \textbf{Avg.} \\ \midrule
\ce{Ca3SbP} & 0.22 & 0.19 & 0.25 & 0.22 & 0.42 & 0.24 & 0.73 & 0.38 \\
\ce{Sr3SbP} & 0.21 & 0.18 & 0.24 & 0.21 & 0.11 & 0.10 & 0.12 & 0.11 \\
\ce{Ba3SbP} & 0.16 & 0.13 & 0.19 & 0.16 & 0.17 & 0.12 & 0.21 & 0.16 \\ \bottomrule
\end{tabular}
\end{table}

\begin{table}
\caption{Hole ($m_\mathrm{h}$) and electron ($m_\mathrm{e}$) effective masses of the orthorhombic \hmn{Pnma} phase of \ce{$A$3SbP}, calculated using HSE+SOC. Effective masses given in units of the electron rest mass, $m_0$}
\begin{tabular}{@{}lcccccccc@{}}
\toprule
 & \multicolumn{4}{c}{$\vb*{m}_\mathbf{h}$} & \multicolumn{4}{c}{$\vb*{m}_\mathbf{e}$} \\ \cmidrule(lr){2-5} \cmidrule(l){6-9} 
\textbf{Compound} & $\mathbf{\Gamma\rightarrow}$\textbf{Z} & $\mathbf{\Gamma\rightarrow}$\textbf{X} & $\mathbf{\Gamma\rightarrow}$\textbf{Y} & \textbf{Avg.} & $\mathbf{\Gamma\rightarrow}$\textbf{Z} & $\mathbf{\Gamma\rightarrow}$\textbf{X} & $\mathbf{\Gamma\rightarrow}$\textbf{Y} & \textbf{Avg.} \\ \midrule
\ce{Ca3SbP} & 0.98 & 0.44 & 0.22 & 0.38 & 0.61 & 0.53 & 1.07 & 0.67 \\
\ce{Sr3SbP} & 1.11 & 0.65 & 0.23 & 0.45 & 0.21 & 0.19 & 0.27 & 0.22 \\
\ce{Ba3SbP} & 0.68 & 0.85 & 1.76 & 0.93 & 0.26 & 0.26 & 0.36 & 0.28 \\ \bottomrule
\end{tabular}
\end{table}

\begin{figure}
\centering
\includegraphics[width=\linewidth]{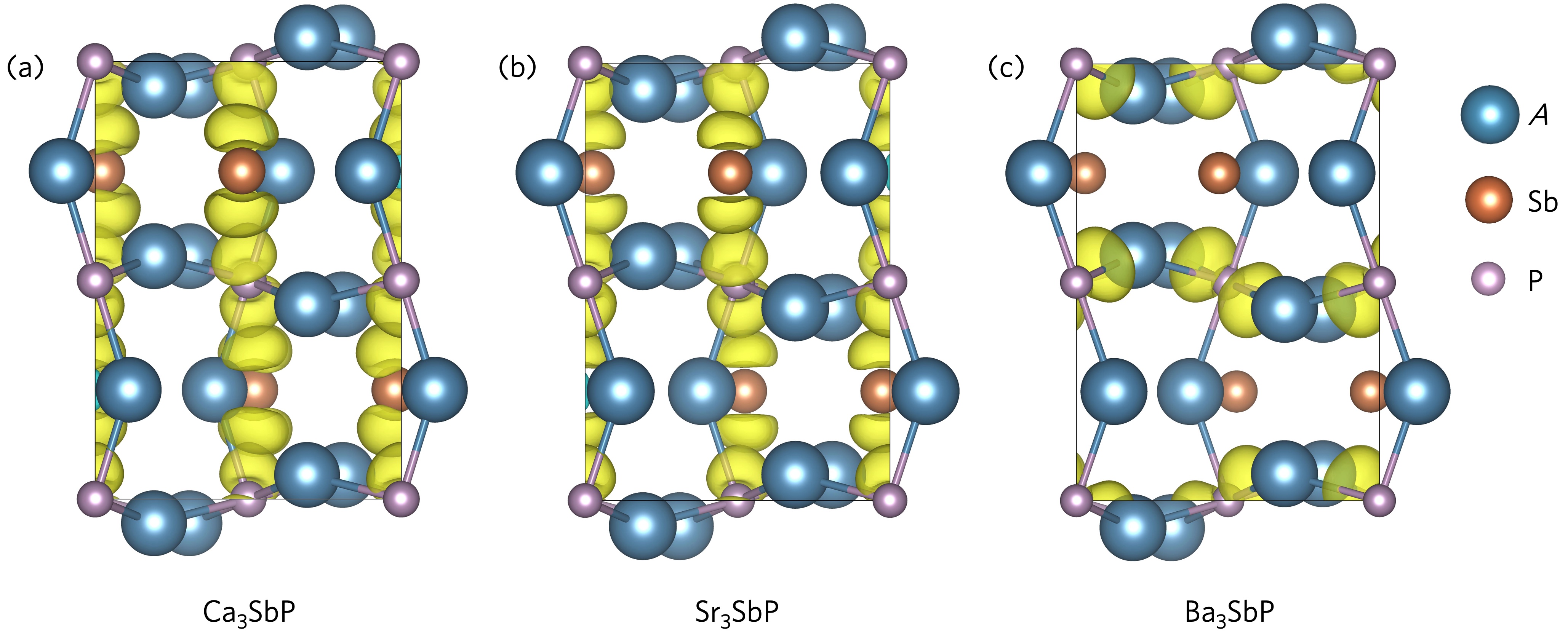}
\caption{Charge density isosurface of the valence band maximum for (a) \ce{Ca3SbP}, (b) \ce{Sr3SbP}, and (c) \ce{Ba3SbP}, calculated using HSE+SOC. The isosurface is displayed at a level of \SI{7}{\meV\per\cubic\angstrom}.}
\end{figure}

\begin{figure}
\centering
\includegraphics[width=\linewidth]{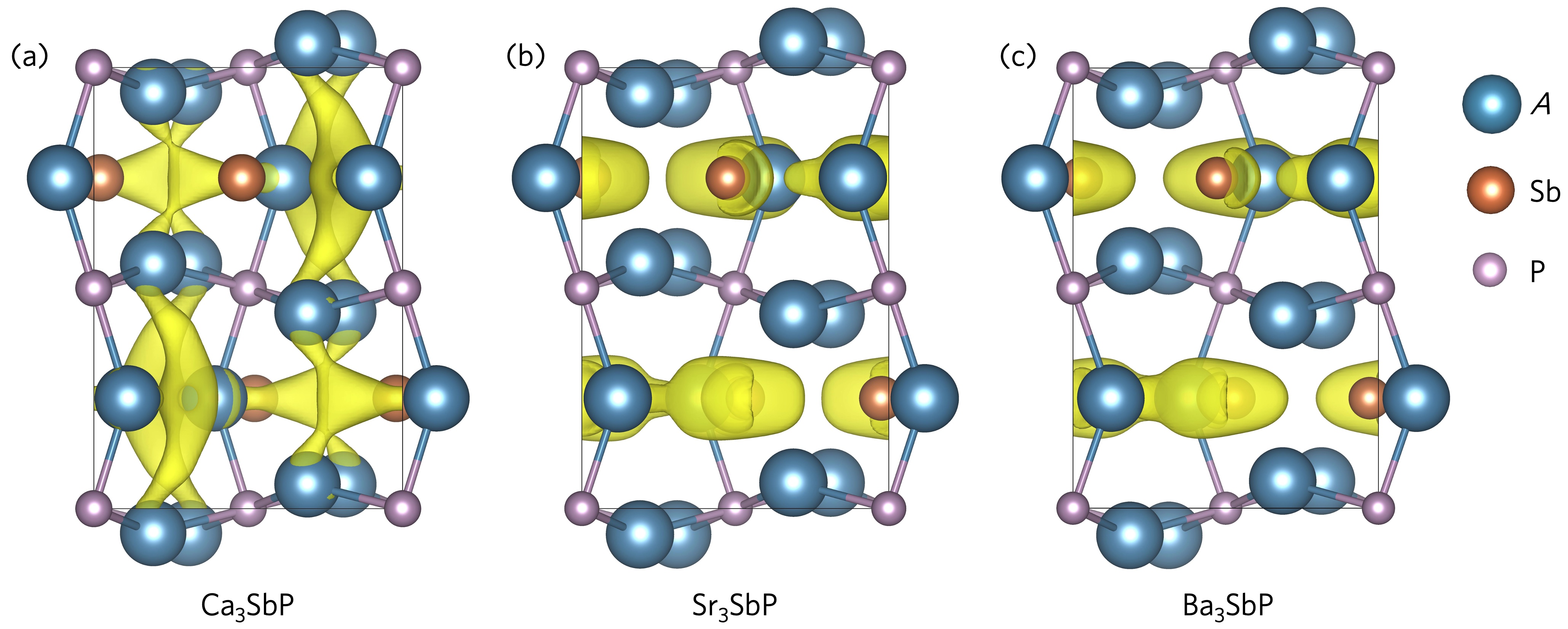}
\caption{Charge density isosurface of the conduction band minimum for (a) \ce{Ca3SbP}, (b) \ce{Sr3SbP}, and (c) \ce{Ba3SbP}, calculated using HSE+SOC. The isosurface is displayed at a level of \SI{5}{\meV\per\cubic\angstrom}.}
\end{figure}

\clearpage
\section{Optical properties}

\begin{table}
\caption{High frequency dielectric constant for the cubic \hmn{Pm-3m} and orthorhombic \hmn{Pnma} phases of \ce{$A$3SbP} calculated using HSE06+SOC}
\begin{tabular}{lcccc}
\toprule
 & \bhmn{Pm-3m} & \multicolumn{3}{c}{\bhmn{Pnma}} \\ \cmidrule(lr){2-2}  \cmidrule(l){3-5}
\textbf{Compound} & \textbf{XX} & \textbf{XX} & \textbf{YY} & \textbf{ZZ} \\ \midrule
\ce{Ba3SbP} & 4.3 & 3.7 & 3.5 & 3.5 \\
\ce{Sr3SbP} & 4.7 & 7.1 & 6.9 & 6.8 \\
\ce{Ca3SbP} & 4.9 & 7.7 & 7.5 & 7.4 \\\bottomrule
\end{tabular}
\end{table}

\begin{figure}
\centering
\includegraphics[width=.7\linewidth]{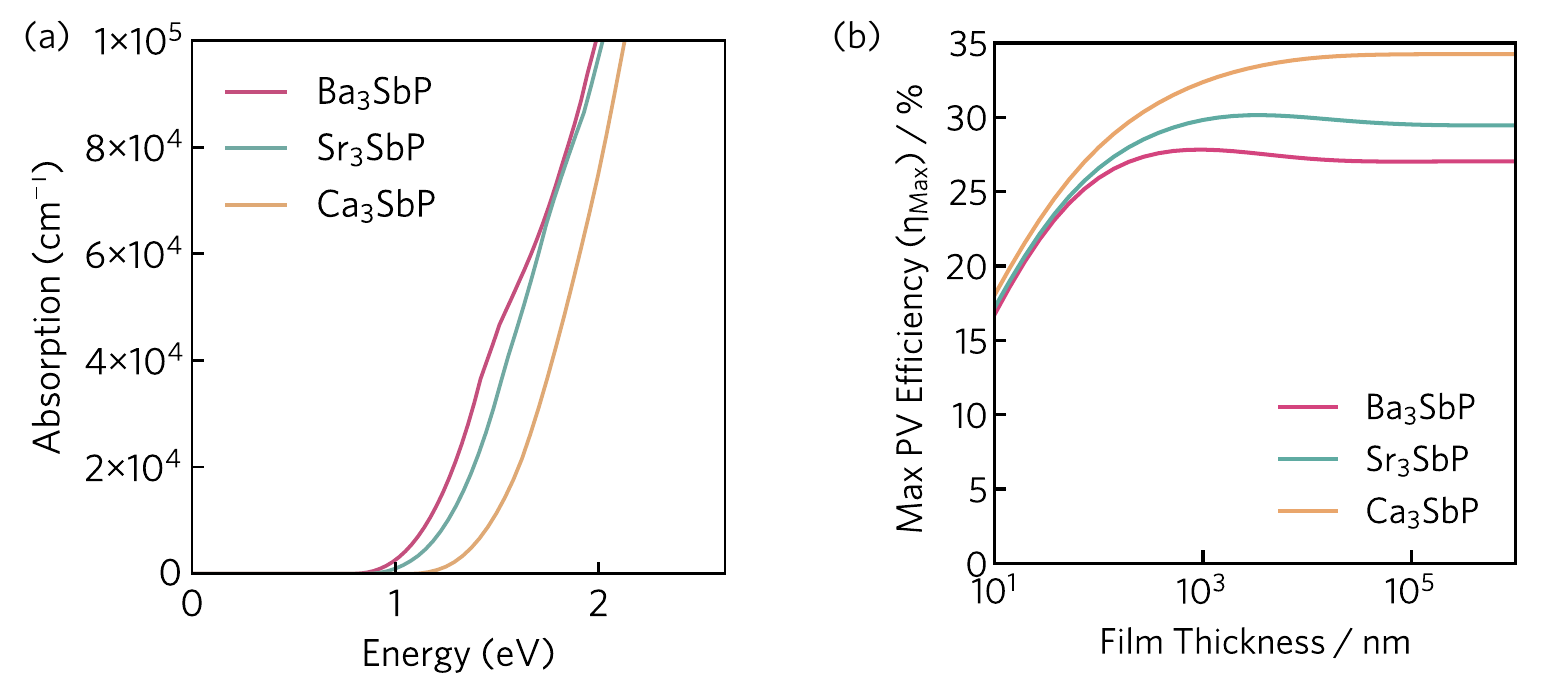}
\caption{(a) Optical absorption calculated using HSE+SOC and (b) thickness-dependent maximum theoretical efficiency calculated using the Blank approach with a Lambertian surface for cubic \hmn{Pm-3m} structured \ce{$A$3SbP}.}
\end{figure}